\documentclass{neutrinos19}

\def\Journal#1#2#3#4{{#1} {\bf #2}, #3 (#4)}

\def\NIMA{{\em Nucl. Instrum. Methods} A}

\def\PRD{{\em Phys. Rev.} D}

\def\be{\begin{equation}}
\def\ee{\end{equation}}
\def\bea{\begin{eqnarray}}
\def\eea{\end{eqnarray}}

\newcommand{\Photo}{}

\begin{document}
\vspace*{4cm}
\title{Status of the Jiangmen Underground Neutrino Observatory}

\author{ Cong Guo, on behalf of the JUNO collaboration }

\address{Institute of High Energy Physics, Chinese Academy of Science, 19B YuQuan Road, Shijingshan District, Beijing, China}

\maketitle\abstracts{
The Jiangmen Underground Neutrino Observatory is a multipurpose neutrino experiment designed to determine neutrino mass hierarchy and precisely measure oscillation parameters by detecting reactor neutrinos from the Yangjiang and Taishan Nuclear Power Plants, observe supernova neutrinos, study the atmospheric, solar neutrinos and geo-neutrinos, and perform exotic searches, with a 20-thousand-ton liquid scintillator detector of unprecedented 3\% energy resolution (at 1 MeV) at 700-meter deep underground. In this proceeding, the subsystems of the experiment, including the cental detector, the online scintillator internal radioactivity investigation system, the PMT, the veto detector, the calibration system and the taishan antineutrino observatory, will be described. The construction is expected to be completed in 2021.}

\section{Introduction}

The Jiangmen Underground Neutrino Observatory (JUNO), a multipurpose neutrino experiment, was proposed for neutrino mass hierarchy (MH) determination by detecting reactor antineutrinos from nuclear power plants (NPPs) as a primary physics goal~\cite{JUNO}. The excellent energy resolution and large fiducial volume anticipated for the JUNO detector offer exciting opportunities for addressing many important topics in neutrinos and astro-physics.

\begin{figure}[htb]
\centering
\includegraphics[height=6.5cm]{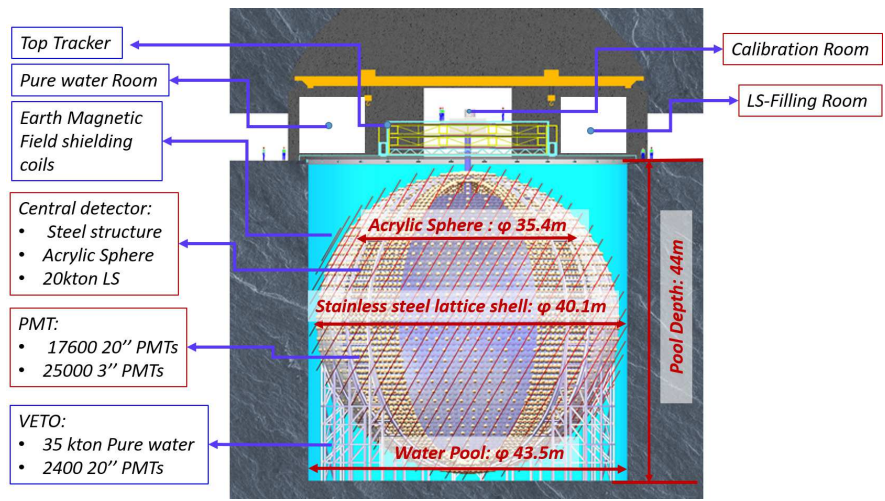}
\caption{Scheme of the JUNO detector. }
\label{fig.JUNO}
\end{figure}

JUNO consists of a central detector (CD) and a veto detector. The central detector, which contains 20 kton liquid scintillator, about 18000 20 inch photomultiplier tubes (PMTs) and 25000 3 inch PMTs, is designed to have a very good energy resolution of 3\%/$\sqrt{E(MeV)}$ and a long lifetime of over 20 years. The veto detector, which consists of a water Cherenkov detector and a muon tracker detector, is used for muon detection as well as muon induced background study and reduction. To supress the radioactivity and cosmogenic background, the CD is submerged in the water Cherenkov detector, which is a pool filled with ultra-pure water and instrumented with about 2000 20 inches microchannel plate photomultiplier tubes (MCP-PMTs). The top tracker (TT) from OPERA experiment~\cite{OPERA1,OPERA2} will be placed on the top of the water Cherenkov detector to behave as a cosmic muon tracker. Four complementary calibration system will be installed to deploy multiple radioactive sources in various locations inside the detector for achieving $<$1\% energy scale uncentainty. The Taishan Antineutrino Observatory (TAO) will be built and operated next to the taishan nuclear power plant to reduce systematic effects in the reactor antineutrino spectrum measured by JUNO.

\section{The JUNO experiment}
The JUNO experiment was proposed in 2008 for neutrino mass hierarchy determination by detecting reactor antineutrinos from nuclear power plants\cite{PRD2009,Yifang,Jun}. The JUNO site is located in Jinji town, Kaiping city, Jiangmen city, Guangdong province, which is optimized to have the best sensitivity for mass hierachy determination and it is at a distance of 53~km from both Yangjiang and Taishan NPPs. The thermal power of the NPPs are expected to be 26.6 GW$_{th}$ at the end of 2020.

Strict requirements for each subsystems have been put forward according to the Monte Carlo simulation results~\cite{JUNOdesign}, in this section, the following subsystems will be described: the CD, the online scintillator internal radioactivity investigation system (OSIRIS), the PMT, the veto detector, the calibration system and the TAO.

\subsection{The cental detector}
The CD of JUNO aims to measure the neutrino energy spectrum using 20 kton LS. An inner sphere with a diameter of about 35.4~m is designed to contain the ultra-pure LS and an outer structure, which is made by stainless steel, with a diameter of about 40.1~m is used to support the inner sphere. The acrylic sphere will be composed by 265 pieces of 120~mm thick spherical panels. The net weight of the acrylic sphere is about 600 tons.

In order to achieve the  3\%/$\sqrt{E(MeV)}$ enrgy resolution, the central detector is required to maximize the collection of optical signals from LS and minimize the background from a variety radioactive sources. The LS recipe, which is optimized from Daya Bay experiment~\cite{DYBLS}, consists of Linear Alkyl Benzene (LAB) as solvent, 2.5 g/L 2,5-diphenyloxazole (PPO) as the fluor and 3 mg/L p-bis-(o-methylstyryl)-benzene (bis-MSB) as the wavelength shifter. The comprehensive light yield is required to be larger than 1200 photon-electrons (P.E.)/MeV and the attenuation length should be longer than 20 m at 430 nm wavelength. The LAB will be filtered with Al$_2$O$_3$ column to remove the impurities in it. Furthermore, a system with functions of distillation, water extraction and steam stripping will be sued to reach the radioactive requirements, namely $^{238}$U$<$10$^{-15}$g/g, $^{232}$Th$<$10$^{-15}$g/g and $^{40}$K$<$10$^{-17}$g/g, respectively~\cite{NIMA,Monica}. A pilot LS purification system has been installed at Daya Bay since 2017~\cite{LS-pilot}.

\subsection{The Online scintillator internal radioactivity investigation system}
\begin{figure}[htb]
\centering
\includegraphics[height=6.5cm]{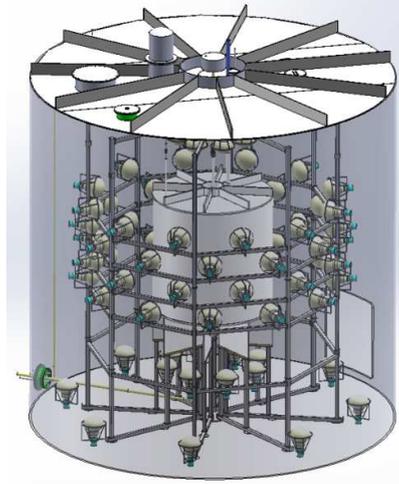}
\caption{Scheme of the OSIRIS facility. }
\label{fig.OSIRIS}
\end{figure}

The radiopurity of the LS will be essential for the success of the JUNO experiment. In order to provide the radio-purity data during the commissioning of the purification system and to ensure the flawless operation by continuously monitoring the LS radio-purity during the CD filling, the OSIRIS system will be built and installed close to the JUNO detector. Fig.~\ref{fig.OSIRIS} shows the scheme of the facility. The tested LS will be contained in the acrylic cylinder, which is located in the center with a dimension of 3~m in diameter and 3~m in height.  81 20 inch PMTs will be installed around the acrylic cylinder to detect the photons. In order to reduce the radio-activities, the acrylic cylinder will be submerged in ultra-pure water and the water tank will also be instrumented with 12 20 inch PMTs to behave as a water Cherenkov detector. The system dimensions are optimized to reach a sensitivity to the baseline radio-purity of 10$^{-16}$g/g for U/Th within 1 day measurement for about 19 tons LS.

\subsection{PMT}
Photon detectors measure the scintillation light created by the interactions of the neutrinos with LS and are key components for accomplishing the physical goals of JUNO. Two sizes of PMTs will be used in JUNO, namely 20000 20 inch PMTs and 25000 3 inch PMTs. About 18000 20 inch PMTs, 13000 MCP-PMTs from North Night Vision Technology (NNVT) and 5000 R12860HQE dynode PMTs from Hamamatsu, will fully cover the CD and contribute about 75\% photocathode coverage. The JUNO collaboration has built two systems to measure the performance of these tubes, which has been running since July 2017~\cite{PMT} and more than 10000 tubes have been tested already. The testing results show that the NNVT PMTs have lower after pulse probability and lower radioactive background, while the Hamamatsu PMTs are better at transit time spread. In order to prevent chain reaction caused by single PMT imploding, an acrylic protective cover will be placed on the top of each 20 inch PMT.

In order to improve the energy scale precision, in particular, the coupling of non-linearity and non-uniformity, 25000 3 inch PMTs, which contribute about 2.5\% to the photocathode coverage, will be installed. The 3 inch PMTs almost always work at the SPE mode for electron anti-neutrino events and are expected to have almost zero dynamic range, hence virtually no non-linearity, thus providing a linear reference to 20 inch PMTs. All the 3 inch PMTs will be supplied by Hainan Zhanchuang Company (HZC) and almost 15000 have been tested and accepted.

\subsection{The veto detector}
In order to reduce the experimental background, the neutrino detector must be placed in deep underground and a veto system will be used to tag muons. The JUNO veto detector consists of a top tracker detector and a water Cherenkov detector.

\subsubsection{The top tracker detector}
\begin{figure}[htb]
\centering
\includegraphics[width=6.5cm]{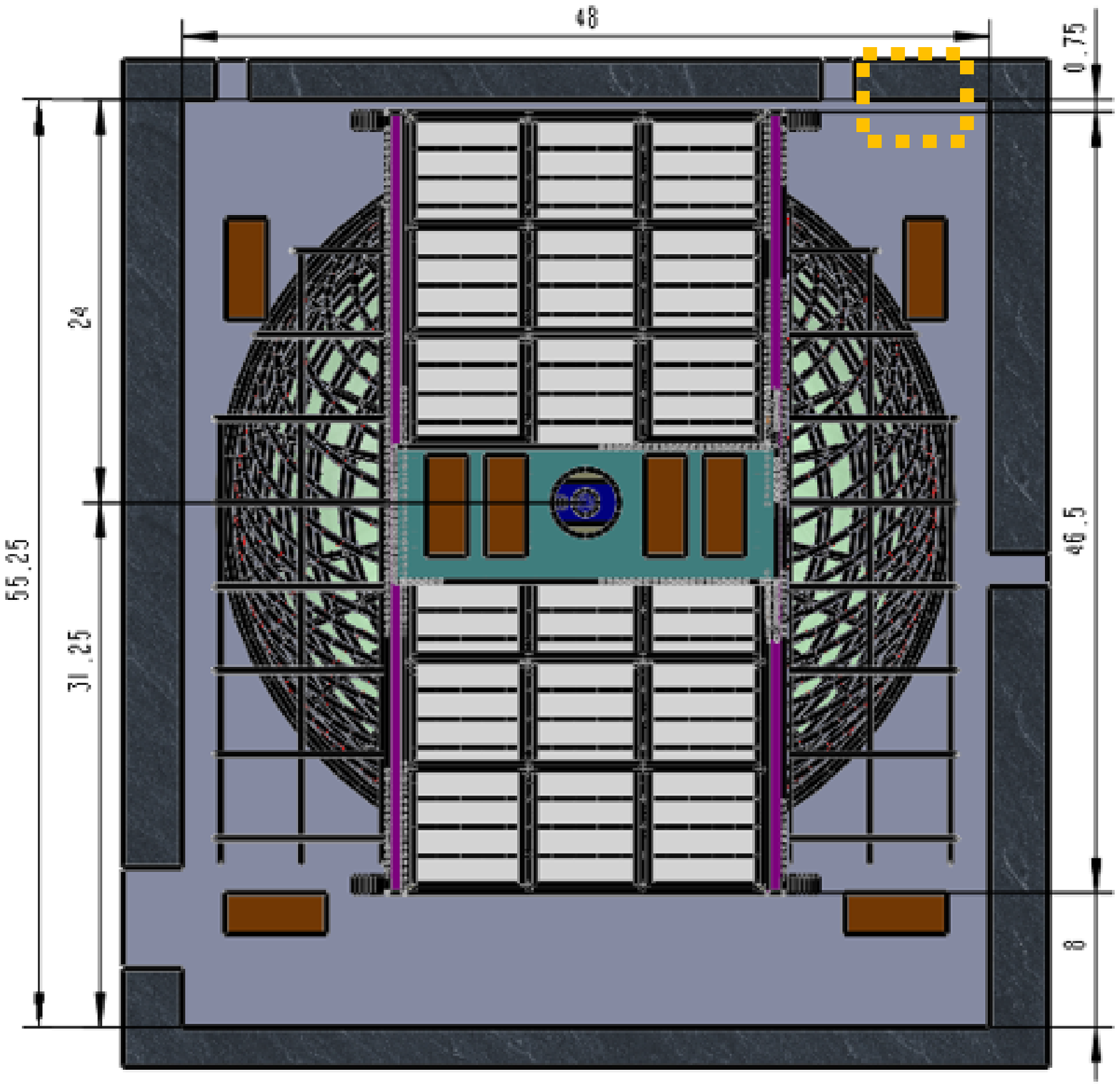}
\includegraphics[width=9cm]{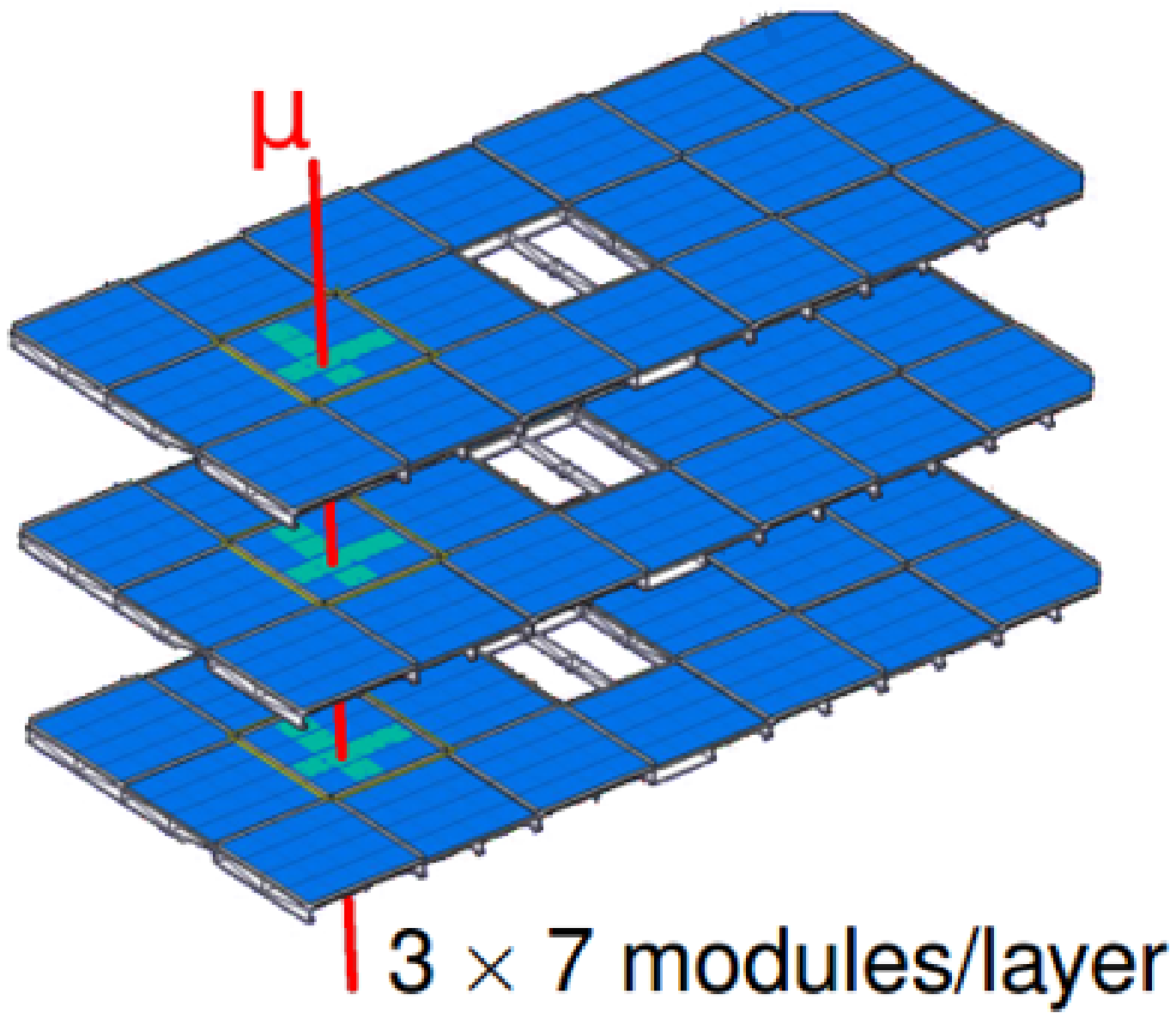}
\caption{Left: Top view of the detector. Right: The arrangement of the TT.}
\label{fig.TT}
\end{figure}

The top tracker (TT) will precisely measure the muon tracking and provide valuable information for cosmic muon induced background study. JUNO TT, which are reusing the OPERA target tracker~\cite{OPERA2}, is made of crossing planes of plastic strips. As the left picture of fig.~\ref{fig.TT} shows, the TT covers more than 60\% of the water pool top area and it can detect $\sim$ 1/3 of muons crossing the CD. The 62 square walls, with a dimension of 6.8 $\times$ 6.8 m$^2$ for each, will be arranged in a 3 $\times$ 7 grid on 3 layers, which is shown in the right picture of fig.~\ref{fig.TT}. All the plastic modules are already at JUNO site and no significant aging has been observed.

\subsubsection{The water Cherenkov detector}
The water Cherenkov detector is a cylindrically shaped tank filled with $\sim$40 kton ultra-pure water and instrumented with about 2000 MCP-PMTs. The water pool has a dimension of 43.5~m in diameter and 44~m in height and is lined with high density polyethylene plate (HDPE) to prevent the radon emanated from the rock to diffuse into the water. The surface of the detector is covered by tyvek as reflecting and diffusing sheets to increase the light collection.

A reliable water system is essential for JUNO detector operation and three basic requirements has been put forward for the ultra-pure water system:
\begin{description}
\item[\texttt{A}] There will be different kinds of materials submerged in the water, including stainless steel, tyvek, PMTs and cables. The complexity of the underground environment makes it difficult to seal the pool completely and almost impossible to keep the water quality good for a long time. Therefore, the first function of the water system is to keep the water quality good to make sure the attenuation length longer than 30~m while the detector is running~\cite{JUNOdesign}.
\item[\texttt{B}] The temperature stability of the central detector is crucial for the entire experiment, thus one of the most important functions of the water system is to keep the overall detector temperature stable~\cite{JUNOdesign}.
\item[\texttt{C}] Radon is one of the most important background sources of the detector and it can emanate from various radium containing substances. Thus a radon removal system has to be consisted in the water system to keep the radon concentration below 0.2mBq/m$^3$~\cite{Rnremoval}.
\end{description}

\subsection{The calibration system}
\begin{figure}[htb]
\centering
\includegraphics[height=6.5cm]{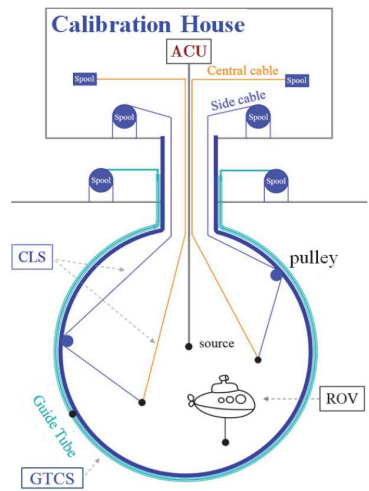}
\caption{Scheme of the calibration system. }
\label{fig.calibration}
\end{figure}

The non-uniformity of the detector response, which is governed by the detector geometry and the optical properties of the relevant detector components, is one of the main contributors for the LS energy resolution. In order to achieve an energy scale uncertainty of better than 1\%, an efficient calibration system is of great importance. As is shown in fig.~\ref{fig.calibration}, the calibration system will consist of four complementary subsystems:
\begin{description}
\item[\texttt{A}] Automated Calibration Unit (ACU). The ACU is a one dimensional system and can be operated along the vertical axis.
\item[\texttt{B}] Cable Loop System (CLS). The CLS is a two dimensional system and can be operated to scan the vertical planes.
\item[\texttt{C}] Guide Tube Calibration System (GTCS). The GTCS is also a two dimensional system and can be operated to scan the outer surface of the CD.
\item[\texttt{D}] Remotely Operated Vehicle (ROV). The ROV is a three dimensional system and can move freely within the CD to fully scan the whole detector.
\end{description}

Three kinds of radiation sources, including neutron sources ($^{241}$Am-Be, $^{241}$Am-$^{13}$C, $^{241}$Pu-$^{13}$C, $^{252}$Cf), positron sources ($^{22}$Na, $^{68}$Ge, $^{40}$K, $^{90}$Sr) and $\gamma$ sources ($^{40}$K, $^{54}$Mn, $^{60}$Co, $^{137}$Cs), will be used for calibration~\cite{calibration}.

\subsection{The Taishan Antineutrino Observatory}
\begin{figure}[htb]
\centering
\includegraphics[height=6cm]{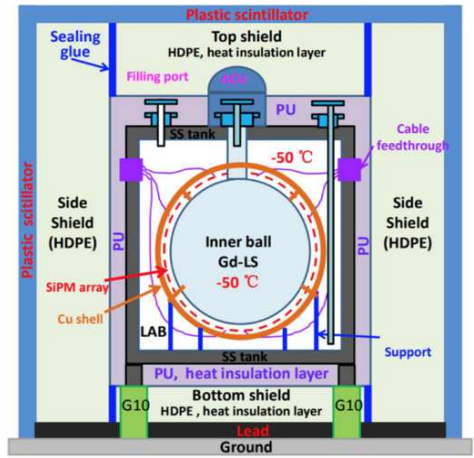}
\caption{Scheme of the TAO. }
\label{fig.TAO}
\end{figure}
The existing model for the energy dependent reactor flux is subject to the anomalous bump observed at 5 MeV and the unknown fine structure of the spectrum~\cite{TAO}. Taishan antineutrino observatory, a ton level Gd-LS detector, will be placed at about 30~m distance to a 4.6~GW$_{th}$-power core of the Taishan NPPs to measure the shape of the un-oscillated reactor antineutrino spectrum to eliminate the possible model dependence for JUNO MH determination. As is shown in fig.~\ref{fig.TAO}, an acrylic sphere, filled with Gd-LS, will be submerged in LAB and silicon photomultipliers (SiPMs), which will be cooled to -50 centigrade and fully cover the detector, will be used for photon detection. The detection efficiency of the SiPM will be $\sim$50\% and a light yield of 4500 photon-electrons at 1 MeV will be reached. The energy resolution will be better than 3\%/$\sqrt{E(MeV)}$~\cite{JUN-TAO}.

\section{Conclusion}
JUNO is a 20 kton LS detector currently under construction in the south of China. The main physical goal is to determine the neutrino MH with six years of running with a significance of 3-4$\sigma$. An unprecedented energy resolution of 3\%/$\sqrt{E(MeV)}$ based on a light yield of 1200~P.E./MeV and an energy scale uncertainty of $<$1\% is required. The civil construction and R$\&$D works for each subsystems are undergoing and data taking is expected to start in 2021.

\end{document}